# Tapered Optical Fiber-based Detection of Charged Particle Irradiation in Space Exploration and Nuclear Reactors

*Manoj K. Rajbhar, Basudeba Maharana, Shyamapada Patra and Shyamal Chatterjee[*]*

*School of Basic Sciences, Indian Institute of Technology Bhubaneswar, Jatni, 752050, India*

[*]Corresponding Author E-mail: shyamal@iitbbs.ac.in

**Abstract:**

In this work, we demonstrate the use of tapered optical fibers (TOF) to detect charged particle (ions), irradiated at various energies, fluences and species. The single mode tapered optical fiber has been used in various sensing applications in recent times. Here, tapered optical fibers have been exposed to two different ion species namely $Ar^+$ and $N^+$ at different energies and different fluences, respectively. The optical spectrum analyser (OSA) detects the changes in the free spectral range (FSR), period, and transmission power loss from the ion beam irradiated TOFs. The change in the refractive index of the cladding material due to the implanted ions influences the transmission spectra of the TOFs and we could able to detect ions of energy as low as 80 keV. COMSOL simulation results are employed to explain the observed changes in spectra. The ion beams induced surface modification and defect formation as well as the implantation in TOF have been predicted using Monte Carlo based 3D TRI3DYN ion-solid interaction simulation and corroborated with other experimental studies such as scanning electron microscopy and Raman scattering spectroscopy. Such tapered optical fiber-based detection technique will help to develop portable device to detect charged particles in space exploration and in nuclear reactors.

Key Words: Tapered Optical Fiber (TOF), Ion beam irradiation, COMSOL, TRI3DYN, Optical Spectrum Analyser (OSA), Ion detection

## 1. Introduction:

Radiation detectors must be light, portable, and highly sensitive to radiation sources that are smaller and more difficult to detect [1]. There are three basic types of radiation detectors: gas-filled detectors, solid-state detectors, and detectors that use scintillators [2]. Ionization chambers, proportion counters, and Geiger-Mueller (G-M) tubes are examples of gas-filled detectors [3]. All of these detecting methods operate on the same concept; the key difference is the voltage supplied across the detectors. Gas-filled detectors are ionisation chambers of ion chambers that register only primary ions at lower voltages (in actuality pair of ions created: a positively charged ion and a free electron). However, one significant disadvantage of such gas ionisation chambers is that they are unable to distinguish between different types of radiation and their energy levels [4]. This is overcome by proportional counters, which are highly useful for spectroscopy since they react differently at different energies, allowing them to distinguish between different types of radiation [5,6]. The second major radiation detection technology is based on solid-state or semiconducting materials and employs a p-n junction [7]. The radiation traverses the depletion area, which generates an electron-hole pair and an analogue signal [8]. In a position sensitive detector which is one kind of photodetector, the energy and position of a light spot can be determined after signal processing. The scintillation detectors are the type of detectors that detects the amount of light produced in some specific crystalline materials when exposed to ion irradiation [9]. Due to its great effectiveness in detecting charge particles and photons, modern scintillation detectors based on optical fiber have efficiently and reliably worked in many high-energy physics experiments [10] [11]. Scintillating fibers are widely used as detectors to measure beam luminosity and charge particle time flight separation. The fiber structure enables efficient light collection and transmission via total internal reflection in the cladding region. The transmission loss during light propagation along the fiber surface is caused by Rayleigh scattering, which is caused by inhomogeneities, radiation damage, and self-absorption [12]. The upgradation of large hadron collision beauty spectrometer is planned to replace the silicon microstrip and gas drift tubes detector by scintillating fiber of diameter 250 µm due to their high homogeneity and low materials cost [13]. Radiation tests by the fiber carried out on the different particles (proton, gamma ray, X-ray irradiation) of different energies show that the expected yield loss is nearly about 40 percentage. The major features of the detector measured by the 2.4 m long SCSF-78MJ fiber using pion/proton beams at CERN and electron beam at DESY shows the detection efficiency 99 percentage with spatial resolution 70 µm [13]. The effects of irradiating the fiber with the same fluence at high power

(short time) or low power (long time) are not well understood. The effect of irradiation fluence rate on fiber and their recovery after irradiation has been investigated in some research [14–20]. According to these studies, radiation damage in optical fibers increases with decreasing radiation fluences. The recovery time of an optical fiber detector is nearly 100 times faster than the recovery time of a scintillator and fiber detector. The optical fibers based on the new Nanostructured Organosilicon Luminophores (NOL11 & NOL19) have a high photoluminescence quantum light yield and a very short delay time (1.34ns and 1.18 ns) [21].

In recent years, optical fiber based techniques in conjunction with nanotechnology have emerged as successful and efficient platform for developing sensing devices [22,23]. Sensors based on tapered optical fiber (TOF) have versatile performance. Due to its small size and the enhanced evanescent field distribution at the tapered region, the TOFs have been implemented for detection of molecules and nanoparticles [24], [25]. Optical fibers are used in several studies to develop radiation sensors. Depending on operating principles, there are several well-known methods used for radiation sensing. For instance, detection of radioluminescence, Cerenkov radiation or analysis of change of refractive index of optical fibers and optical absorption analysis due to radiation are used for detection of electrons and gamma rays. Other popular methods involve scintillating materials, thermoluminescence and optically simulated luminescence materials. Some of the most used techniques include refractive index sensing [26], Rayleigh scattering loss/transmission monitoring [27], fluorescence labelling [28], intra-fiber modal interference [29], Raman spectroscopy-based sensor [30], chemiluminescence [31], surface plasmon resonance [32], Fiber Bragg Gratings (FBG) [33]. Lee *et al.,* developed an optical fiber-based dosimetry for detection of electron irradiation using Cherenkov radiation method [34]. Arvidsson *et al.,* measured ionizing fluence of photon irradiation using hard and multimode index multimode fibers [35]. A real time scintillating fiber based dosimeter was developed by Bartesaghi *et al.,* for measurement of fluences of neutron and gamma ray during radio therapy [36]. Brichard *et al* worked on improving radiation hardening of optical fibers used in plasma diagnostics [37]. Alfeeli *et al.,* embedded scintillating material within a holey optical fiber structure and used Cherenkov radiation to detect gamma rays [38]. However, use of tapered optical fibers to detect heavier charged particles is handful at this stage.

In this work, we show ion irradiation detection capacity of tapered optical fiber (TOF) by wavelength interrogation technique. The TOF has been irradiated with nitrogen and argon ions at two different energies of 80 and 100 keV and at different fluences, respectively. The comparison of optical spectra from pristine and the irradiated fibers are clear indicatives of implantations and irradiation induced defects in the fibers. The ion implantations and induced defects have been predicted by Monte Carlo based simulation of 3D structures of the fiber and corroborated with scanning electron microscopy and Raman scattering studies respectively. We have used COMSOL simulation to predict the change of optical spectra due to irradiations.

## 2. Principles:

### 2.1 Tapered Fiber:

The tapered fiber, which is made by drawing the fiber uniformly when it is heated to a certain temperature to reduce its diameter to less than 10 µm. It has three regions; untapered region, which retains the diameter of the original fiber taken, the conical taper transition region with gradually changing diameter, and taper waist segment, which has a very small and uniform diameter. Due to heating and drawing, at the taper waist region the core and cladding diameters decrease resulting in much smaller core and cladding regions (shown in the supplementary figure S1). Due to lower core region, the guided modes redistribute resulting in excitation of higher order modes along with fundamental mode. The higher order modes of TOF have greater overlap over the cladding region that allow for interaction of light with the cladding region. Moreover, these modes have different effective indices and the difference between these indices can be represented as $\Delta n$. These modes overlap and interfere about the waist region such that the transmitted intensity can be given as:

$$I = I_1 + I_2 + 2\sqrt{I_1 I_2} \cdot \cos(\Delta\varphi)$$

Where, $I_1$ and $I_2$ are the intensities of $HE_{11}$ and $HE_{12}$ mode, respectively and the phase shift: $\Delta\varphi = 2\pi \frac{\Delta n.L}{\lambda}$, here, L is the length of taper waist, $\lambda$ is the wavelength of light. The condition for constructive interference is $\Delta\varphi = 2m\pi$, where m is an integer. So, the wavelength value about the interference peak can be written as:

$$\lambda_m = \frac{\Delta n.L}{m}$$

The fringe spacing of the interference pattern termed as free spectral range (FSR) can be expressed as $FSR = \frac{\lambda^2}{\Delta n.L}$ that depends on length of the waist region. The fringe spacing can be varied by varying the taper waist length.

An important parameter in any modal interferometer is visibility for different sensing applications. Higher visibility is needed to get an accurate measurement as the signal-to-noise ratio will be more. The general expression for V is: $V = \frac{I_{max}-I_{min}}{I_{max}+I_{min}}$, which can be simplified as defined as [39]:

$$V = \frac{2\sqrt{k}}{1+k}$$

Here, $k = \frac{I_1}{I_2}$. In this work, we have expressed visibility in terms of fringe contrast (FC) (in dBm) that can be defined as $FC = -10\log(1-V)$.

**2.2 Ion Beam Irradiation:**

Ion beam irradiation is a very well-developed field of study for materials modification, analysis, in semiconductor industry and for radiation therapy. The space is full of such particle radiation, which is harmful for living beings and equipment for space explorations. Furthermore, there are irradiations build up in nuclear reactors. These charged particles or ions interact with matter via nuclear and electronic stopping interactions. Such interactions lead to surface modifications, defect generations and thermal spike in a material. The consequence may lead to change of crystal phase, surface roughening or patterning, atomic mixing, change of chemical state, mechanical and structural changes to name a few. Usually, low energy ions lose energy in solid via nuclear stopping and high energy ions lose through electronic stopping interactions. Nuclear stopping causes displacement of atoms in a solid and creates defects apart from implantation of the ions. Electronic stopping causes excitation and ionization of electrons in a solid leading to generation of thermal spike and lattice heating. In this study the aforementioned ion irradiation effects in tapered optical fibers are exploited to detect the irradiation.

3. **Experimental Details:**
3.1 **Fabrication of Taper Fiber:**

Single-mode fibers (SMF-28) with core/cladding diameters of 8/125 μm were used in this study. The protective coating was removed from a 5 mm long section of the SMF along the length. The TOF was fabricated by flame and brush technique in which the fiber section is heated to nearly its melting point and pulled from both sides to cause lowering of diameter. The diameter of waist region and the length of the microfiber was kept at 10 μm and 7.5 mm, respectively [40]. Multiple such TOFs were fabricated and their transmission spectrum was recorded using spectrum analyser before and after the irradiation. The fringe spacing and visibility parameters of each of the TOF was evaluated.

### 3.2 Ion Beam Irradiation:

The irradiation was done with $Ar^+$ and $N^+$ ions at various energies ranging from 80-100 keV. For all the ion energies, the irradiation was done at fluences of $3\times10^{16}$ and $5\times10^{16}$ ions.cm$^{-2}$ respectively. The fiber is placed on the aluminium substrate surfaces on which a predefined cylindrical hole of 10 mm radius has been made to irradiate the tapered region of fiber uniformly from both sides. This aluminium substrate holding tapered fiber kept perpendicular to the incoming beam in all the implantations. $Ar^+$ and $N^+$ ion implantations of energies 80 and 100 keV have been done in the low energy ion beam facility (LEIBF) of IUAC, New Delhi.

### 4. Results and Discussion:
### 4.1 FESEM Analysis:

Surface of tapered part of the fiber was studied using Scanning Electron Microscope (Merlin Compact with GEMINI-I column, Zeiss Pvt. Ltd, Germany) for both the pristine and the irradiated fibers [41]. From the Fig. 1, it is evident that irradiated surface is more roughened up than the pristine fiber. The magnified image in the inset suggests surface roughening of the surface. Such surface roughening is associated with ion beam sputtering, and surface diffusion. In earlier studies on oxide, it was noted that due to preferential sputtering oxygen vacancies and surface defects are found. Such modified surface by the ions results in change of the refractive index of the cladding and thereby allowing more modes in the cladding region than in the core [42][43].

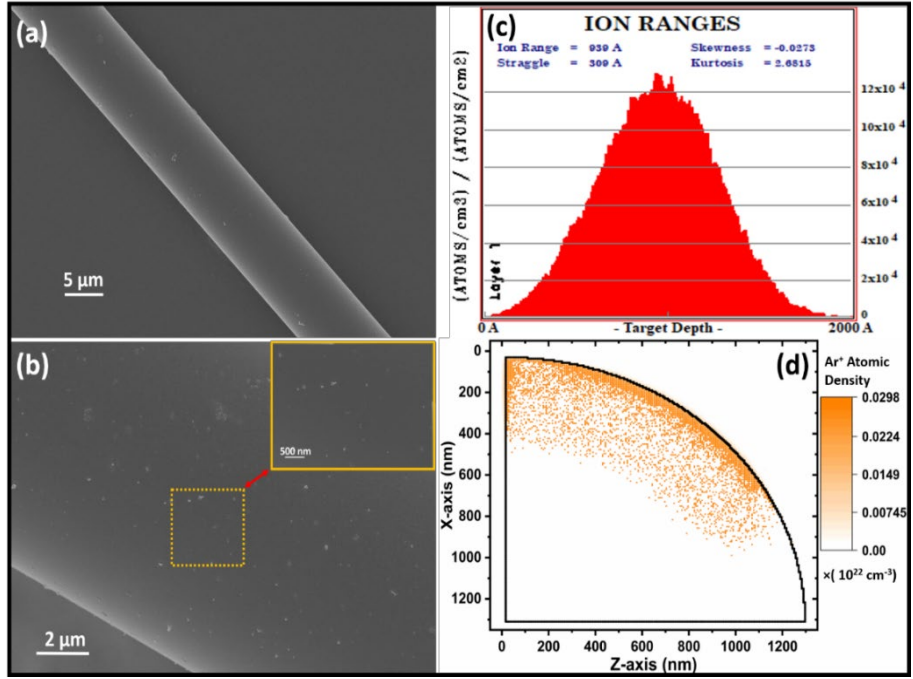

***Figure 1:*** *FESEM image of Pristine (a) and Irradiated (b) tapered fiber region with $Ar^+$ ion at 100 keV energy and $5\times10^{16}$ $cm^{-2}$ fluence. (c) SRIM calculation showing ion range of $Ar^+$ in tapered microfiber. (d) TRI3DYN simulation of irradiated $Ar^+$ ions over quarter part of microfiber and atomic densities distribution of $Ar^+$ ions after irradiation of 100 keV and at an ion fluence of $5\times 10^{16}$ $cm^{-2}$.*

### 4.2 Raman Spectra Analysis:

Raman scattering spectroscopy is used to study surface defects produced by ion irradiations. Fig. 2 shows the Raman scattering spectrum of the pristine fiber and the irradiated fiber. The figure further depicts a comparative study after irradiating with $N^+$ and $Ar^+$ ions at 100 keV energy and at a fluence of $5x10^{16}$ $cm^{-2}$. The peaks at about 493 $cm^{-1}$, 606 $cm^{-1}$, and 795 $cm^{-1}$ are attributed to $SiO_2$ of the optical fiber [44]. After irradiating with $N^+$ the peak at 794.83 $cm^{-1}$ disappears, which may be due to ion induced defects formed on the surface. For the case of $Ar^+$ irradiation, a peak shift of about 3 $cm^{-1}$ is observed which might be due to collision with comparatively heavier ions, causing significant number of defects as evident from simulations presented later in this manuscript. We further observe appearance of a small peak around 795 $cm^{-1}$. These results show that we the surface cladding material is modified with ion irradiations for both light and heavy ion species.

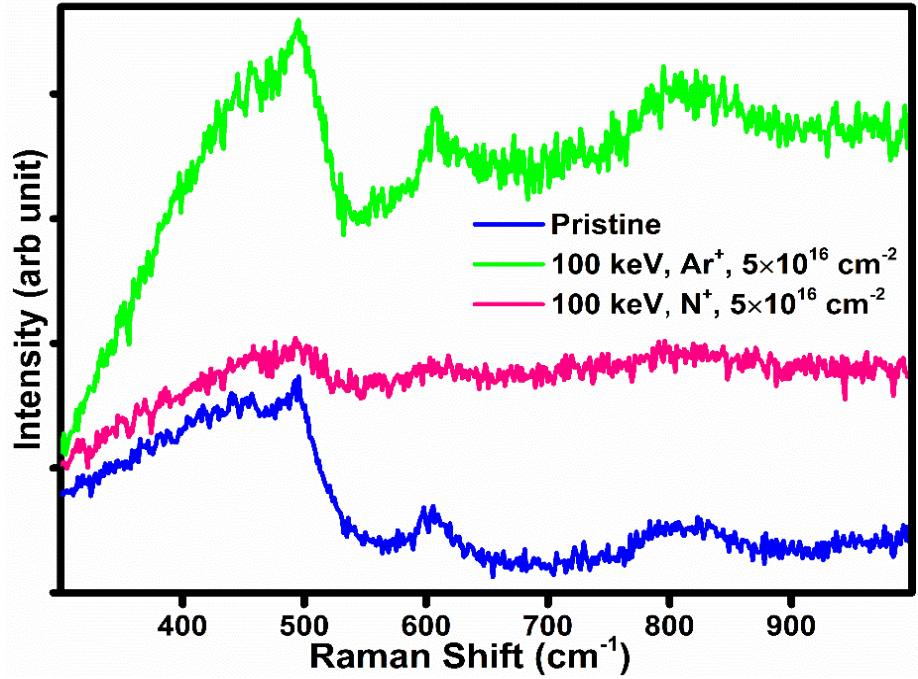

***Figure 2:*** *Raman scattering spectra of pristine, Ar⁺ irradiated, and N⁺ irradiated optical fiber.*

### 4.3 OSA (Optical Spectrum Analyser) Data Analysis:

Figure 3 shows the transmission spectra of ion beam irradiated and the pristine tapered optical fibers. Figure 3(a) represents the spectrum of a pristine TOF and TOF irradiated with nitrogen ions at two different energies of 80 and 100 keV for the same fluence of $5 \times 10^{16} cm^{-2}$. Similarly, Fig. 3(b) gives a comparison of two spectra from pristine and argon ion irradiated TOFs at two different energies and at same fluence. It is observed that as the energy of the ions increases, the visibility parameter decreases. The cause is attributed to the increase in penetration depth of ions in cladding region which increases its effective index. This leads to spreading of core mode over the cladding region that lowers the $I_1$ value and decreases visibility parameter. The contrast range of transmission was between -18 dBm to -28 dBm. Also, the free spectral range (FSR) gradually increases as we increase the energy of the ions. Due to increase in energy of ions, the ions get implanted deeper in cladding region and the effective index of the cladding region increases that decreases *Δn* value. As the argon-ion is heavier compared to nitrogen ion, the modification is more prominent at the same energy and fluence for argon ion irradiation (as shown in the table 1). Hence the transmission power loss for Ar⁺ is more than that of N⁺.

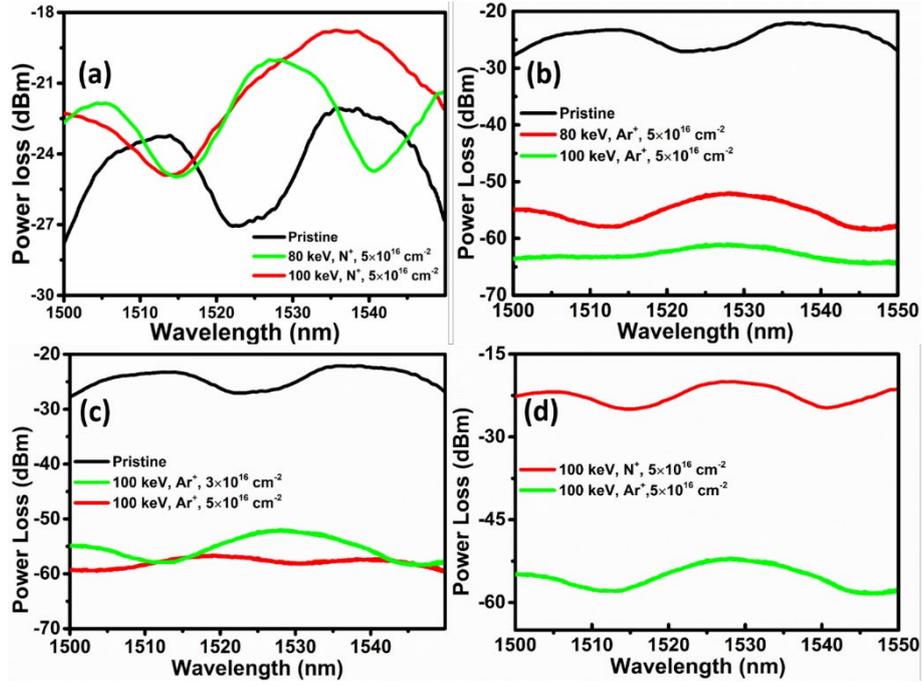

***Figure 3:*** *Transmission spectrum of (a) $N^+$ ion irradiated at different energies for a fluence of $5\times 10^{16} cm^{-2}$, (b) $Ar^+$ ion irradiated with different energies at fluence $5\times 10^{16} cm^{-2}$ (c) $Ar^+$ irradiated with the same energy at 100 keV with different fluence, (d) $N^+$ vs. $Ar^+$ with the same energy and same fluence.*

**Table 1.** SRIM data of damages event on optical microfiber by both $Ar^+$ and $N^+$ ions of energy 100 keV.

| Damage Event | $Ar^+$ ion irradiation | $N^+$ ion irradiation |
|---|---|---|
| Ion Ranges in the Sample | 93.9 nm | 147.7 nm |
| Total Displacement of atom | 1344/ion | 462/ion |
| Total Vacancies | 1307/ion | 449/ion |
| Si vacancies | 0.6/ion | 0.16/ion |
| O - vacancies | 0.6/ion | 0.14/ion |
| Sputtering Yield | 2.434/ion | 0.479/ion |
| Energy Loss | 106 eV/nm | 67.7 eV/nm |

Figure 3(c) shows the spectra of Ar+ ions irradiated at different fluences and at the same energy. As the fluence increases, the refractive index of the cladding increases due to a greater number of ion implantations and defects [42][43]. So, the effective index of the cladding region increases and decreases *Δn* value. This increases the FSR while lowering the visibility of spectra.

Figure 4(a) depicts the variation of FSR with energy of ions for $Ar^+$ and $N^+$. The FSR changes for increase in energy of the ion at constant value of fluence. Additionally, the change in FSR is more prominent for argon ions irradiated TOFs than that for nitrogen ions irradiated TOFs. Figure 4(b). shows the spatial frequency spectra corresponding to the transmission spectra of pristine TOF, $Ar^+$ irradiated, and $N^+$ irradiated optical TOFs. From this graph, it is evident that the interference spectra are due to overlap of two prominent modes. The modes are distinct for the pristine TOF. For the irradiated TOFs, the intensity of the fundamental mode decreases significantly. This is due to increase in effective cladding index by ions implantations that lowers the mode confinement in diminished core region. The fundamental mode then spreads across the cladding region that lowers its intensity.

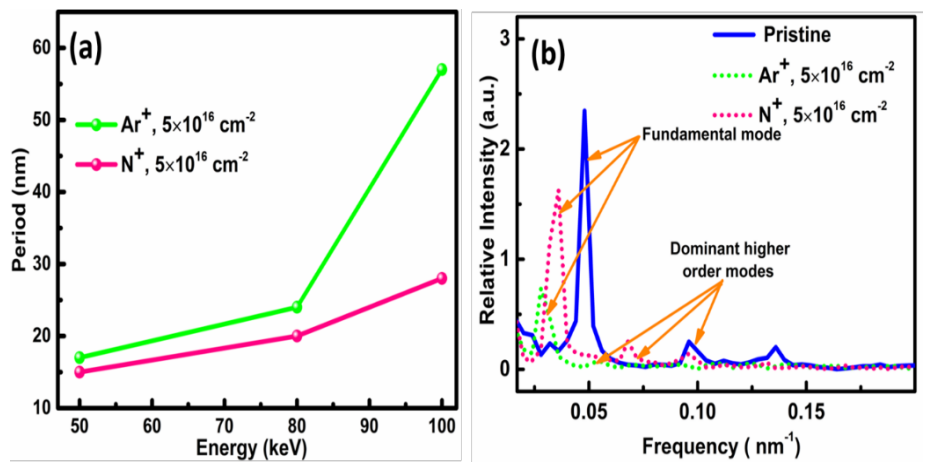

***Figure 4:*** *(a) Variation of period at different ion beam energy, and (b) the corresponding spatial frequency spectrum of the transmission spectrum of the pristine, $Ar^+$ irradiated, and $N^+$ irradiated.*

### 4.4 TRI3DYN Simulation:

A TRI3DYN computer simulation program used in this study, predicts the interaction of ion beam irradiation into any arbitrary 3D nanostructure system under the concept Monte Carlo simulation [45,46]. This simulation is an advanced version of well known TRIDYN simulation, which is able to explain the dynamic collision simulation only in one direction. In the case of TRI3DYN simulation, the entire computational volume is divided into equisize small volumes which are commonly known as voxels. Each voxel is defined with some local atomic density. Due to ion beam irradiation, the local atomic density of voxels is changed due to the presence of foreign atoms having different atomic densities and recoil relocation relaxations. Ion beam irradiation influences the one voxel to interact with another, also responsible for the transportation of nanostructure materials towards or away from the nanostructure surface. Initially, in this Simulation, we first defined the arbitrary shape and element distribution taking part in the nanostructure, then we chose irradiation conditions with a wide range of beam profilers. A state-of-the-art TRI3DYN computer simulation introspects the detailed potential artefact with the variation of composition profile, surface sputtering, and surface contour. The TRI3DYN Simulation is mainly based on the binary collision approximation (BCA), which is the utmost treasure tool for the accurate and fast prediction in the various ion-solid interaction processes. Monte Carlo Simulation or Monte Carlo method was used to carry out the simulations. BCA, due to its linear cascade model, is highly capable of giving satisfactory precision for crystalline as well as amorphous materials at smaller computational expenses while evaluating the ion range and distribution of ion-induced damage. The TRI3DYN simulation can work in both static and dynamic mode of variables of the solid target and it takes into account sputter and recoil atoms, target atomic re-deposition, and atomic mixing [47]. The program is able to predict defect migrations, distribution of defects and implanted ions, relative sputtering yield, and re-deposition of sputtered-out atoms to name a few and predictions have been corroborated well with different experiments [48–52]. TRIM simulation, which is the part of the freely available ion solid interaction simulation package SRIM, is mainly using the BCA code and predicts the outcome for amorphous materials after ion beam interaction and to describe the sputtering after TRIM.SP simulation is used, which is a derived version of TRIM. In TRI3DYN simulation, we subdivided the entire target volume into small in-depth slabs, which represent the fixed atomic volume after relaxing the complete 3D nano-system. The compiled result of several BCA simulations shows the projected range and defect formation by all incident ions such as associated collision cascade, implanted ions, creation of vacancies at the surface by means of sputtering and inside the bulk nanostructure by displacement of the atom.

For the current system, initially, the voxel was arbitrarily loaded with a material atomic component such as Si-O atoms for $SiO_2$ based microfiber and also defined by the surface contour of the structure. In this simulation, each central voxel is surrounded by the 6 first-order neighbour voxels at the surface, 12-second order neighbour at the edge, and 8 third-order neighbour voxels at the corner. The initial atomic density of all components take participate in the simulation is also mentioned. We start our study of the TRI3DYN simulation in case of Quarter part (diameter 2700 nm) and HALF hybrid microfiber (diameter 1350 nm) of $SiO_2$ as shown in the figure 5(a) and supplementry diagram S2 respectively. We choose the full diameter of the quarter part of fiber of $SiO_2$ (Diameter = 2700 nm) and half part of microfiber of diameter 1300 nm respectively. We have computed damage distributions arising from broad beam irradiation for two ions source namely $Ar^+$ ions and $N^+$ ions at the ion energies of 50 keV, 80 keV, and 100 keV and at a fluence of $5\times10^{16}$ ions cm$^{-2}$ using TRI3DYN in static mode. The change in atomic density $SiO_2$ after broad $Ar^+$ ion beam irradiation of 50 keV, 80 keV and 100 keV is shown in the figure 5(b)-5(d) which confirmed that with increase ion energies $SiO_2$ atomic density continue decreasing due to ion irradiation induced defects.

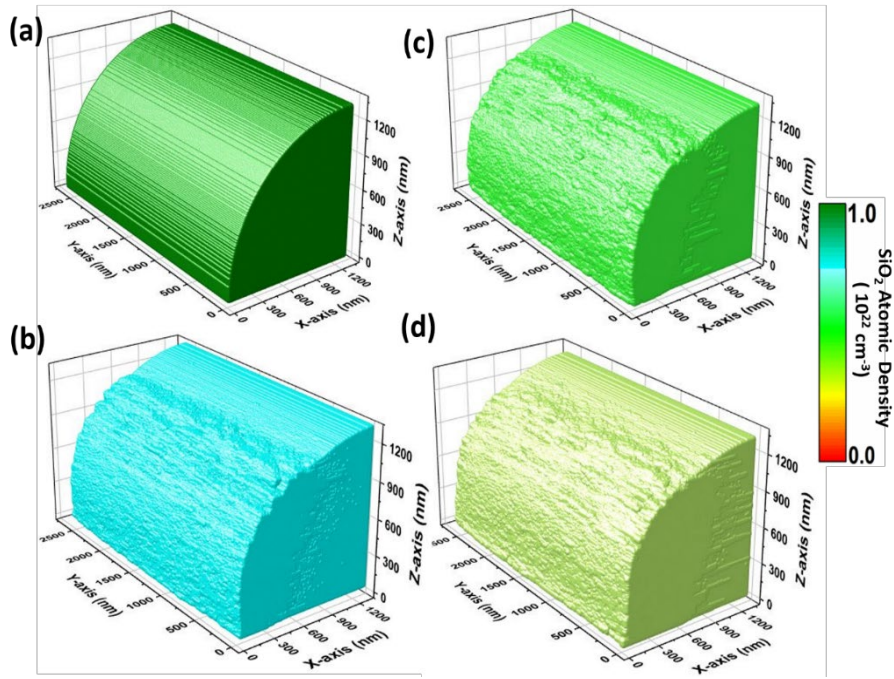

*Figure 5:* TRI3DYN computer simulation model system with a quarter part of 2700 nm diameter of a micro-tapered fiber of length 2800 nm. (a) Part of the fiber before irradiation and after irradiation with $5\times10^{16}$ cm$^{-2}$ by $Ar^+$ ions at an energy of 50 keV (b), 80 keV (c) and 100 keV (d) respectively. The atomic densities are coloured with the scale of $10^{22}$ cm$^{-3}$.

While looking into the atomic fraction of the fiber, it is realized that the collisional effect plays a major role throughout this simulation. The ion irradiation further creates a large number of interstitials, vacancy. This simulation also predicts that with ion beam irradiation at a fluence of $5\times10^{16}$ cm$^{-2}$ there is a change in the atomic fraction of the Si- atom and the O-atom at the surface of the optical microfiber for both $Ar^+$ and $N^+$ ions at various ions with various ion energy range 50-100 keV. The variation of O-atomic fraction and Si-atomic fraction after $Ar^+$ ion irradiation is shown in the figure 6(a)-(f).

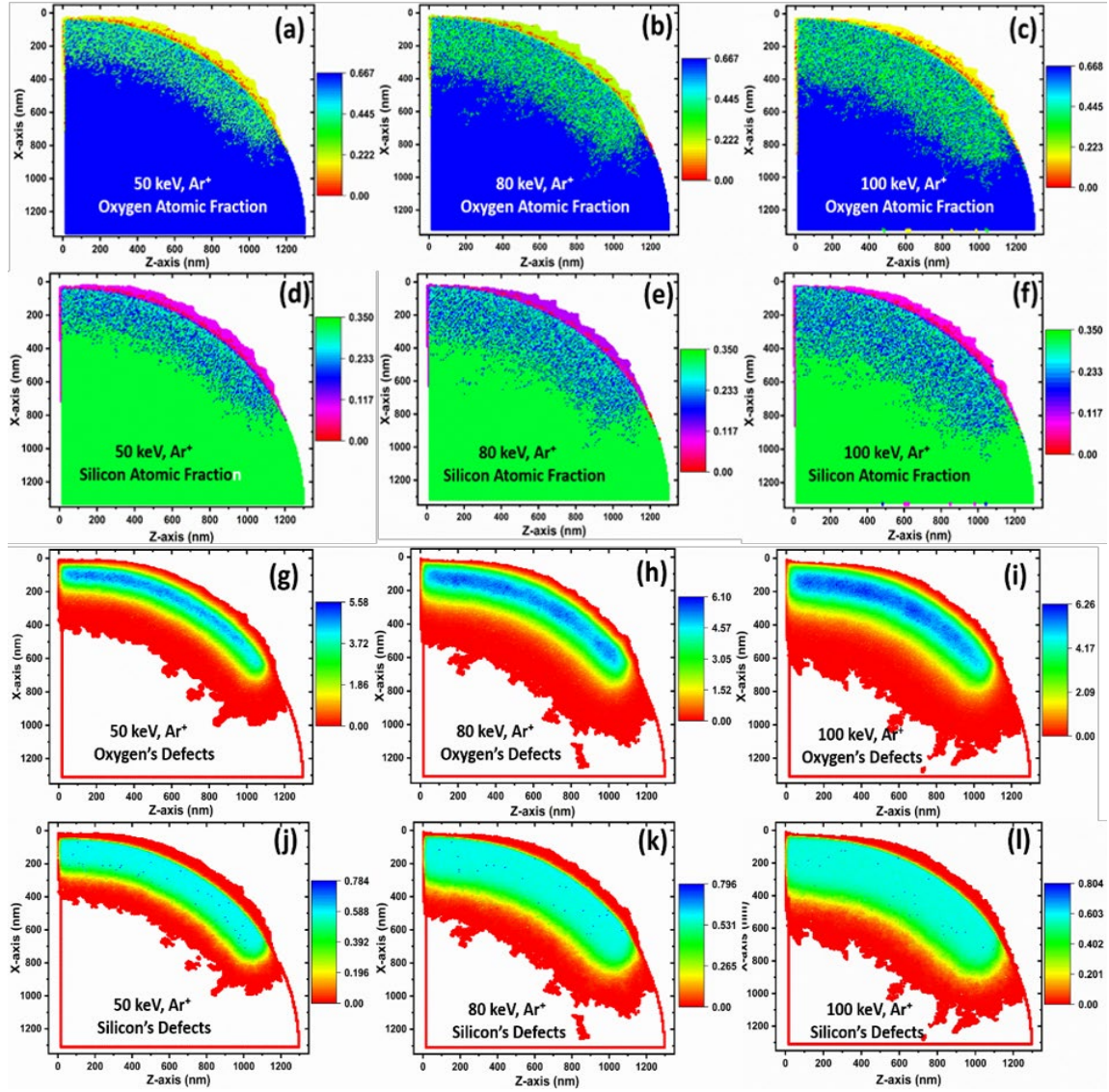

*Figure 6:* TRI3DYN simulation results for change in oxygen atomic densities for 50 keV(a), 80keV (b) and 100 keV (c) , and silicon atomic density for 50 keV (d), 80keV (e) and 100 keV (f) irradiated by $Ar^+$ ions at an ion fluence of $5 \times 10^{16}$ $cm^{-2}$ respectively . Figure 6(g)-(i) shows the change in O-defect with increase ions energy from 50 keV to 80 keV and then 100 keV respectively. Similarly Figure 6(j)-(l) shows the change in Si-defect with increase ions energy from 50 keV to 100 keV followed by 80 keV respectively. All result is 10 nm central slice cut of X-Z plane through the fiber axis. The atomic densities are coloured with the scale of $10^{23}$ $cm^{-3}$.

It is apparent that in case of $Ar^+$ ion irradiation, Si- atomic fraction changes from 0.33 to 0.12 and O-atomic fraction changes from 0.66 to 0.22 at the surface which indicates that silicon atomic fraction changes almost 49% and the oxygen atomic fraction changes almost 65% at the top surface, whereas just little below the surface of the microfiber Si- atomic fraction changes from 0.33 to 0.23 and O-atomic fraction changes from 0.66 to 0.34 indicated that silicon atomic fraction changes almost 30% and the oxygen atomic fraction changes almost 50%. This changes in atomic fraction of the quarter fiber continue increasing in depth with increase in ion energy *i.e.,* 50 keV to 100 keV. The change in atomic fraction arises due to ions induced defect like interstitial and vacancy which may be O-vacancy or interstitial and silicon-vacancy or interstitial. The variation of O-defect (interstitial and vacancy) and Si-defect (interstitial and vacancy) with various energy shown in the figure 6(g)-(l) continue increase in depth with increase in ion energy which confirmed the change in atomic fraction of O-atom and Si-atom at the surface and in the depth of the microfiber. Here, one point must be noted that we get more O-defect which is around 6 dpa (displacement per atom) than Si- defect which is around 0.7 dpa. Similar type of observations we get when we used the half microfiber of diameter 1300 nm (As shown in the Supplementary S3). It was discovered that using TRI3DYN simulation on pristine quarter part of a microfiber of diameter (2700 nm) in static mode is shown in the figure 7(a), and a broad beam irradiation of ions source, $N^+$ ions at energies 50-100 keV and at fluence of $5\times10^{16}$ cm$^{-2}$ resulted in damage distributions in the microfiber. The atomic density of SiO2 decreased with increasing ion energies after exposure to broad $N^+$ ion beams at energy 100 keV, as shown in Figures 7(b).

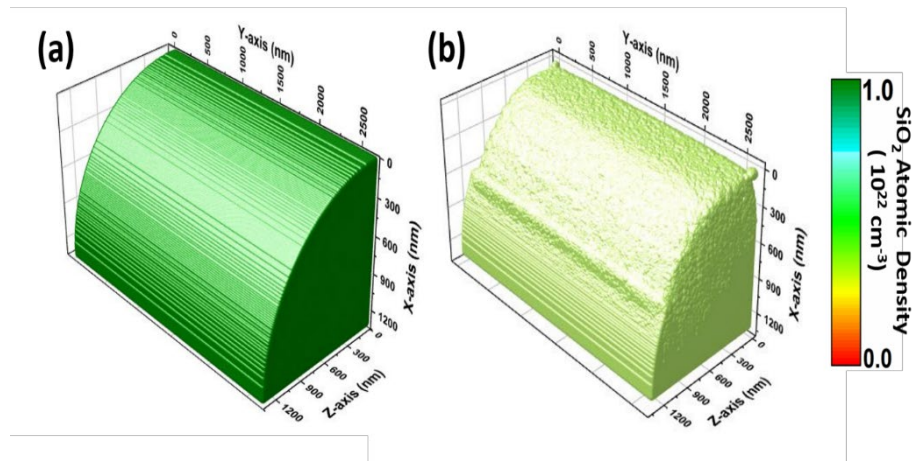

*Figure 7: TRI3DYN computer simulation for quarter part of micro-fiber of diameter 2700 nm. (a) Part of the fiber before irradiation and after irradiation with $5\times10^{16}$ cm$^{-2}$ by $N^+$ ions at an energy 100 keV (b) respectively. The atomic densities are coloured with the scale of $10^{22}$ cm$^{-3}$.*

The variation of O-atomic fraction and Si-atomic fraction after $N^+$ ion irradiation is shown in the figure 8(a)-(f).In case of $N^+$ ion irradiation, Si- atomic fraction changes from 0.33 to 0.2 and O-atomic fraction changes from 0.66 to 0.40 at the surface which indicates that silicon atomic fraction changes almost 39% and the oxygen atomic fraction changes almost 40% at the top surface, whereas just little below the surface of the microfiber Si- atomic fraction changes from 0.33 to 0.23 and O-atomic fraction changes from 0.66 to 0.44 indicated that silicon atomic fraction changes almost 30% and the oxygen atomic fraction changes almost 33% is shown in the figure 8(a)-(f). The variation of O-defect (interstitial and vacancy) and Si-defect (interstitial and vacancy) with various energy shown in the figure 8(g)-(l). The defect depth in case of $N^+$ ions is greater than as compare to $Ar^+$ ions as the mass of the $Ar^+$ is large compare to $N^+$ ions, so its range is less as compare to $Ar^+$ ions. Similar type of observations we get when we used the half microfiber of diameter 1300 nm and irradiated with $N^+$ ions (As shown in the Supplementary S4) The percentage of changes in the O-atomic fraction is larger as the compared Si-atomic fraction due to the large preferential sputtering of O-atom. There are further implantations of argon atoms close to the surface after irradiation. Since the change in atomic densities of the medium affects its refractive index, hence this influences the light traveling through that medium.

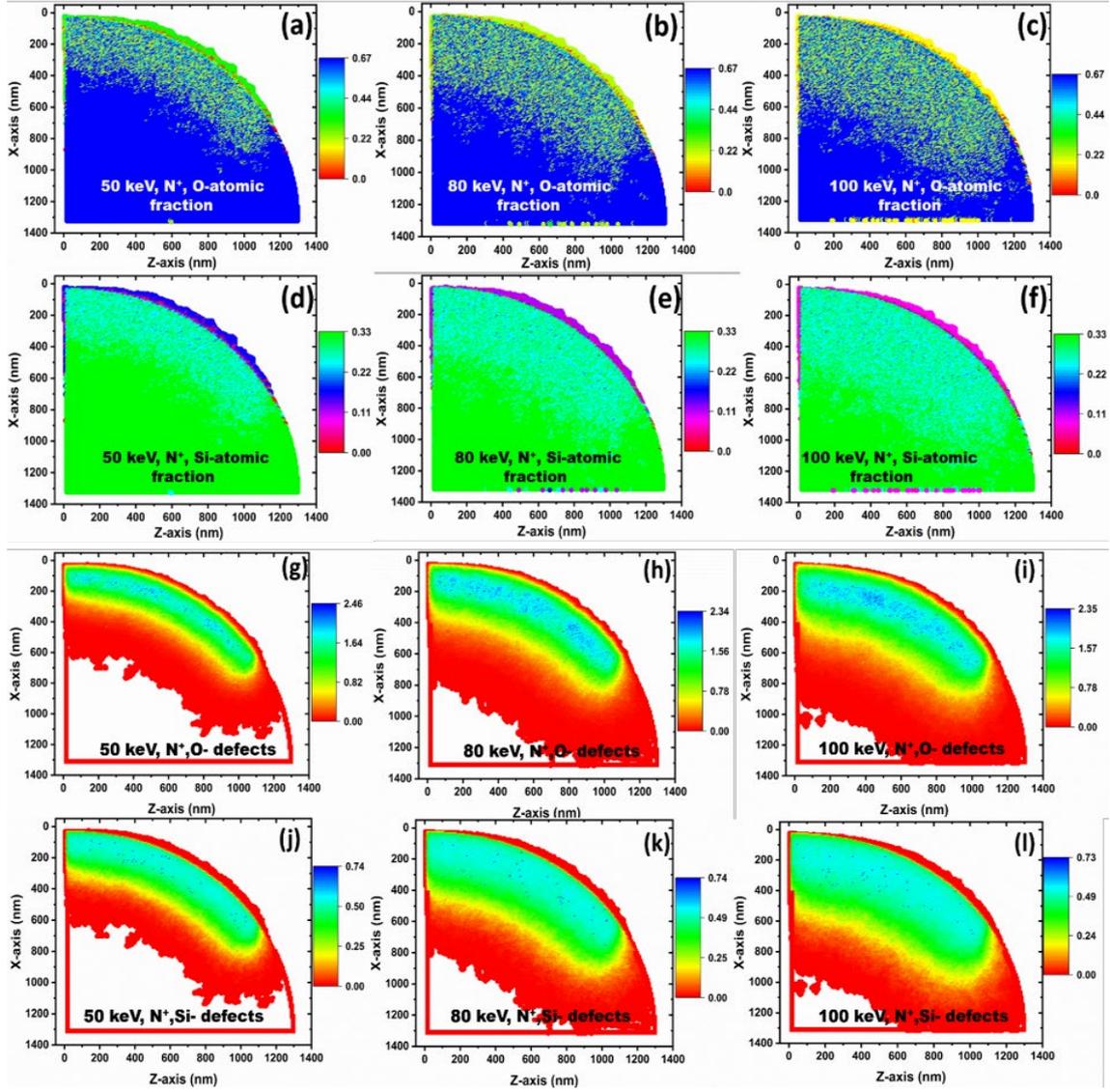

*Figure 8:* TRI3DYN simulation results for change in oxygen atomic densities for 50 keV(a), 80keV (b) and 100 keV (c) , and silicon atomic density for 50 keV (d), 80keV (e) and 100 keV (f) irradiated by $N^+$ ions at an ion fluence of $5 \times 10^{16}$ $cm^{-2}$ respectively . Figure 6(g)-(i) shows the change in O-defect with increase ions energy from 50 keV to 80 keV and then 100 keV respectively. Similarly Figure 6(j)-(l) shows the change in Si-defect with increase ions energy from 50 keV to 100 keV followed by 80 keV respectively. All result is 10 nm central slice cut of X-Z plane through the fiber axis. The atomic densities are coloured with the scale of $10^{23}$ $cm^{-3}$.

## 5. Conclusions:

Our study shows that tapered optical fibers offer a unique capability to detect the various types of ion irradiations ranging from low energy to fast moving ions. The irradiated ions produce a defect zone and implantation of foreign atoms into the cladding region of tapered section, which changes the refractive index of the cladding medium. As a result, the free spectral range, phase and power loss spectra are quite different for irradiated fibers than the pristine one. With increasing fluence of ions more changes of RI of cladding are expected and as evident from OSA measurements. Furthermore, the higher energy ions penetrate more depth and alter significant thickness of the cladding, which is also reflected in power loss distribution pattern. Thus, we have demonstrated that tapered optical fiber can directly detect ions of various species at wide energy range and for various fluences. More research however, is required for charged particle selectivity and exact energy and fluence calibrations using such system. Such system could be extremely useful for making dosimeter and direct detection of charged particle in space exploration, accelerators and in nuclear power reactors.

**Acknowledgments**

We would like to acknowledge access to the Inter University Accelerator Centre (IUAC, New Delhi) for the Argon and Nitrogen ion-irradiations and Dr. Rajan Jha, Kalipada Chatterjee and Subrat Sahu of IIT Bhubaneswar for OSA facilities. FESEM facility of IIT Bhubaneswar is gratefully acknowledged. The authors are also grateful to Prof. Wolfhard Möller for the support with the TRI3DYN simulation. B.M would like to acknowledge UGC for a fellowship.